# A unifying framework for amyloid-mediated membrane damage: The lipid-chaperon hypothesis

Carmelo Tempra[1,♦], Federica Scollo[2,♦], Martina Pannuzzo[3,♦], Fabio Lolicato[4,*] and Carmelo La Rosa[5,*]

[1]Institute of Organic Chemistry and Biochemistry, Prague, Czech Republic
[2]J. Heyrovský Institute of Physical Chemistry, Czech Academy of Science, Prague, Czech Republic
[3]Laboratory of Nanotechnology for Precision Medicine, Fondazione Istituto Italiano di Tecnologia, Via Morego 30, Genoa 16163, Italy
[4]Heidelberg University Biochemistry Center, Heidelberg, Germany
[5]Dipartimento di Scienze Chimiche, Università degli Studi di Catania, Viale A. Doria 6 – 95125 Catania, Italy

♦The authors contributed equally to this work

*Corresponding authors; fabio.lolicato@bzh.uni-heidelberg.de, clarosa@unict.it

## Abstract

Over the past thirty years, researchers have highlighted the role played by a class of proteins or polypeptides that forms pathogenic amyloid aggregates in vivo, including i) the amyloid Aβ peptide, which is known to form senile plaques in Alzheimer's disease; ii) α-synuclein, responsible for Lewy body formation in Parkinson's disease and iii) IAPP, which is the protein component of type 2 diabetes-associated islet amyloids. These proteins, known as intrinsically disordered proteins (IDPs), are present as highly dynamic conformational ensembles.

IDPs can partially (mis) fold into (dys) functional conformations and accumulate as amyloid aggregates upon interaction with other cytosolic partners such as proteins or lipid membranes. In addition, an increasing number of reports link the toxicity of amyloid proteins to their harmful effects on membrane integrity. Still, the molecular mechanism underlying the amyloidogenic proteins transfer from the aqueous environment to the hydrocarbon core of the membrane is poorly understood.

This review starts with a historical overview of the toxicity models of amyloidogenic proteins to contextualize the more recent lipid-chaperone hypothesis. Then, we report the early molecular-level events in the aggregation and ion-channel pore formation of Aβ, IAPP, and α-synuclein interacting with model membranes, emphasizing the complexity of these processes due to their different spatial-temporal resolutions. Next, we underline the need for a combined experimental and computational approach, focusing on the strengths and weaknesses of the most commonly used techniques. Finally, the last two chapters highlight the crucial role of lipid-protein complexes as molecular switches among ion-channel-like formation, detergent-like, and fibril formation mechanisms and their implication in fighting amyloidogenic diseases.

1.0 **Introduction**

Intrinsically disordered proteins (IDPs)[1] in the aqueous phase are characterized by a not well-defined secondary and tertiary structure since their backbone explores a large number of conformations[2]. Some of these proteins are called metamorphic or chameleonic[3,4], due to the possibility that identical amino acid sequences can adopt either α-helix, β-sheet, or random coil secondary structures. IDPs perform important biological functions, such as cellular signaling and regulation[5]. Amylin (IAPP), amyloid-beta (Aβ), and α-synuclein (αS) proteins linked with type 2 diabetes (T2D), Alzheimer's (AD), and Parkinson's (PD) diseases, respectively, show both IDPs and chameleonic nature. Human IAPP is a 37-amino acids hormone containing a disulphide bridge[6] co-secreted with insulin (ratio 1:100) from islet β-cells. In physiological conditions, insulin, together with glucagon, regulates the level of glucose in the blood. In diabetes type 2, amyloid deposition of IAPP has been associated with islet β-cells death[7]. Aβ is a 39-43 amino acids peptide derived from the cleavage by β and γ secretase of the amyloid precursor protein (APP)[8–11]. Aβ performs several physiological functions, including the regulation of synaptic plasticity, the facilitation of neuronal growth and survival[4,12,13]. The αS protein senses membrane curvature and contributes to synaptic trafficking, vesicle budding, and regulation of dopamine release[14,15]. In pathologic conditions, the accumulation of Aβ and αS is toxic to neurons and the mechanism behind has been associated with membrane disruption and ion dysregulation[16]. Hence, the study of IDPs-membrane interaction appears to be crucial to understand the molecular toxicity mechanism. However, due to the great complexity belonging to cell membranes, a bottom-up approach is strongly needed. Therefore, model membranes are useful to unravel the basic interaction between their constituents and other chemical entities in the aqueous phase. For this purpose, large unilamellar vesicles (LUVs) are the most popular and well-characterized model membranes built at the different complexity levels of composition, to mimic various cell membranes and organelle as well as drug delivery systems[7,17–19].

Although there is a large consensus that toxicity originates from membrane damage, a general molecular mechanism remains unsettled. In this framework, some hypotheses have been proposed: i) the amyloid hypothesis considering fibril structures responsible for membrane damage[20,21]; ii) the toxic oligomer hypothesis in which the toxic species are small unstructured prefibrillar amyloidogenic proteins[22], able to damage the membrane in two independent steps: transmembrane ion-channel-like-pores and detergent-like mechanism[20,21,23,24] iii) the lipid-chaperone hypothesis in which free lipids in the aqueous phase form a stable lipid-amyloidogenic protein complex highly prone to insert into the bilayer[23–29]. The latter hypothesis assumes specific relevance considering that an abnormal accumulation of amyloidogenic peptides, together with a dysregulated phospholipase activity, high levels of unsaturated or short acyl-chains phospholipids, are found in individuals who develop T2D, AD, and PD[30–34]. Moreover, it is crucial to highlight that the hypothesis mentioned above shall not be considered exclusive but rather a solid bridge between the other two. The present review discusses the toxicity models of amyloidogenic proteins focusing on our recently developed lipid-chaperone hypothesis. We identified a common mechanism shared by all the amyloidogenic and non-amyloidogenic proteins, i.e, human and rat-IAPP (h-IAPP and r-IAPP), $A\beta_{(1-40)}$ and α and β-synuclein (αS and βS). In addition, the review describes the fundamental steps that led us to formulate this theory and the need for a multidisciplinary approach to the study of amyloidogenic IDPs and their relation with T2D, AD, and PD diseases.

2.0 **Toxicity models of amyloidogenic proteins**

In 1984, George Glenner purified the Aβ peptide from cerebrovascular amyloid fibrils associated with AD[35]. In the early 1990s, three different mutations were discovered in the genes encoding for the *APP*[36], and for the secretases (prenisilin 1 and 2)[37,38], associated with the onset of the familiar form of AD. Both the observations of Aβ being the main component of plaques and the identification of such mutations favoring the aggregation of Aβ in familiar AD led to the formulation of the "amyloid hypothesis" according to which the accumulation of this peptide could cause the disease. Similarly, abnormal protein deposits arranged in a β-pleated sheet structure were also shown to progressively affect organ or tissue functions in other pathologies such as type II Diabetes, Parkinson's, Huntington's diseases[39].

Above a certain critical concentration, amyloid proteins, despite having a different amino acid sequence, share the characteristic of aggregating into oligomers with common immunological epitopes in different disorders, followed by protofibrils, mature fibrils and finally plaques rich in highly ordered cross-β structures[40].

Initial studies suggested that amyloid fibrils were responsible for cellular toxicity in many different amyloid-type disorders. In cultured neurons, injected Aβ fibrils were shown to affect action potentials, synaptic transmission, and membrane depolarization causing cell death. Likewise, amylin fibrils were toxic to islet cells in vitro. The addition of Congo red was shown to reduce amyloid cellular toxicity by binding to fibrils or inhibiting fibrils polymerization[41,42]. Cell death via apoptosis was also demonstrated after exposing cell cultures to fibrils of αS, the major component of inclusions in PD[43]. Since some amyloid types were found to accumulate only inside the cell and others in the extracellular space, amyloid toxicity was initially postulated to be dependent on the interaction of fibrils with the cell membrane, which faces both compartments. Several experimental studies showed that fibrils, once adhered to the membrane, could cause lipid depletion, induce membrane thinning and consequent disruption, or provoke a strong membrane curvature stress and deformation[44–47]. Despite strong evidence that amyloid fibrils were capable of disrupting membrane integrity, the level of insoluble amyloid aggregates was shown not always to correlate with the severity of clinical symptoms[48–50], suggesting a no one to one correlation between the progress of the disease and the amyloid deposition. The severity of cognitive impairment in AD was later shown to better correlate with low-molecular weight and soluble Aβ aggregates[51]. Similarly, the evidence of cytotoxicity associated with pre-fibrillar amyloid assemblies began to emerge for other peptides associated with amyloid diseases. More specifically, fibrils could be found in only 90% of patients developing T2D[52–54]. The use of rifampicin to prevent hIAPP fibril formation, but not the oligomerization, was proven to be insufficient in protecting beta-cells from apoptosis[55]. Another study on the cytotoxic potential of amylin proved that mature fibrils were not toxic to β-cells[56]. All this together with several other pieces of evidence gave rise to the formulation of the so-called toxic-oligomer hypothesis[57,58], associating the oligomers or pre-fibrils aggregates with the most toxic molecular entities, although a new controversy emerged about the role played by insoluble fibrils. Indeed, these could play a protective role in the sequestration of toxic oligomers or represent a reservoir for their release. To date, three mechanistic models have been proposed to explain the cytotoxicity of prefibrillar assemblies: the "Carpet-Model", "Detergent-like" and transmembrane "channel-like pore". The "Carpet-model" refers to small prefibrillar species binding to the membrane and generating an asymmetric pressure on the two layers of the membrane[59]. The relaxation of this pressure results in the leakage of the membrane. The "Detergent-like" mechanism is a particular case of the previous one. Prefibrillar structures bind to the membrane and a detergent-like activity causes the removal of lipids from one side of the bilayer, generating an asymmetric pressure[59]. The latter mechanism is associated with small protein oligomers that can insert into the bilayer and make a pore similar to a non-specific ion-channel, leading to ionic dyshomeostasis and eventually to cell death. This mechanism has been observed for many amyloidogenic proteins[60–62]. Soluble oligomers of all types of amyloids indeed share common epitopes despite differences in the primary sequence, suggesting that toxicity correlates with their common structure. Experimental studies show that toxicity is strictly related to the exposure of hydrophobic regions of the oligomer[63], which likely mediates its insertion into the cell membrane.

Experimental studies have also evidenced that membrane-embedded oligomers do not always form channel-like structures but can facilitate the transport of molecules across the membrane by inducing membrane packing defects[64]. Moreover, oligomers may also alter the function of transmembrane receptors, as in the case of the N-methyl-D-aspartate receptor involved in the glutamatergic synaptic transmission in AD[65,66].

Considering the wide structural heterogeneity of amyloid assemblies, it is plausible to hypothesize that aberrant interactions can occur on multiple levels. For example, it has been experimentally demonstrated that both Aβ oligomers and fibers contribute to membrane disruption[25]. The two steps-mechanism foresees that upon interaction with the amyloidogenic protein the membrane undergoes irreversible damage due to the formation of small pores, most likely due to toxic oligomers, as a first step. The second step follows, in which the larger aggregates are considered to be responsible for the removal of lipids from the bilayer, i.e., the aforementioned detergent-like mechanism. However, the boundaries or the relationship between the 'toxic oligomer" and 'amyloid' hypotheses remain poorly defined. Nevertheless, the interaction of amyloid peptides with the membrane is widely recognized as their common basis. Although most of the scientific community nowadays supports the 'toxic oligomer hypothesis', this is not comprehensive. It cannot provide all the answers we are currently looking for. Additional complications arise from the complexity of the systems and the experimental limitations. The events taking place in biological processes are challenging to study due to their extremely broad temporal and spatial scales. The challenge becomes harder for IDPs and even more complex for amyloidogenic IDP due to the absence of a fixed secondary structure and their ability to change their secondary structure based on the binding partner. The fast dynamic nature of these proteins makes them very hard to study and characterize.

### 3.0 Synergism between simulations and experiments

Local changes of protein loops and side chains take place in less than a second ($10^{-14}$ - $10^{-1}$ seconds). Rigid body movements, such as subunit and domain adjustments, occur instead in the range of microseconds and seconds, whereas protein folding and oligomerization take place in $10^{-14}$ - $10^{-1}$ seconds. To capture all of these movements ultrafast techniques able to explore a portion of space in the range of 0.01 –10 ångström are needed, but also techniques capable of looking at large-scale behavior in the range of seconds. Unfortunately, right now there are neither experimental nor computational techniques able to achieve, at the same time, the needed spatial and temporal resolutions. Furthermore, high-resolution measurements of molecular structures are only possible for extremely rigid systems, and the analysis of the atom-scale interaction energies involved in these processes are difficult to quantify experimentally. However, the contribution to the development and the better understanding of the amyloid field is undeniably due to the employment of many biophysical techniques, other than the variety of evidence obtained *in vivo* systems. The current knowledge and the progress in the field are due to the great effort made by the scientific community during the past decades, coming especially from the combination of a large variety of biophysical approaches. Indeed, when choosing the right approach, one should bear in mind not only the type of information provided by the technique itself but also the advantages and the limitations of each of those. In this chapter we summarized the strengths and limitations of the most used experimental and computational techniques for studying amyloid IDPs and their interaction with model membranes.

### 3.1 Strength and weaknesses of experiments

The structural characterization of IDPs and the study of their conformational features and changes have aroused enormous interest, although it has been proven to be challenging due to their heterogeneous and dynamic nature[67]. From a historical point of view, Circular Dichroism (CD) and Fourier Transform Infrared Spectroscopy (FT-IR) should be highlighted as classical tools to study indirectly the protein (un)folding and the aggregation processes, by monitoring the change in the secondary structure[68–70]. Random-coil features usually, but not necessarily, dominate the CD spectrum characteristic of IDPs, and amyloid fibers are typically constituted by cross-β sheet structures. Although a large variety of algorithms exist, estimating the exact content of the different secondary structures from a CD spectrum with high precision is not yet a concrete possibility. In contrast, FT-IR spectroscopy provides a better deconvolution and quantification of the various structural components[71]. Attenuated Total Internal Reflection (ATR), for instance, requires micrograms of protein and it is one of the few techniques allowing the study of dry proteins, i.e., protein films[71]. Vibrational spectroscopies have been proven to be particularly useful in monitoring characteristic β-sheet structure[72], although the spectral interferences due to endogenous biomolecules and water represent a substantial issue. To this respect, isotopic labeling and unnatural amino acids approaches enable to improve the sensitivity for the secondary structure and to detect residue-specific information of the surrounding polarity[73–75]. Besides that, the X-ray scattering, Electron Spin Resonance (ESR), and Nuclear Magnetic Resonance (NMR) provide insights into the ensemble of conformations sampled by disordered states[76,77]. Particularly, NMR related approaches, including heteronuclear NOE (Nuclear Overhauser Effect) measurements, NMR coupling constants, residual dipolar couplings (RDCs), NMR line splitting, pulsed-field gradient NMR, and heteronuclear single quantum coherence (HSQC), are considered to be extremely informative, because more specific and qualitative compared to the previous ones. Those are used to prove local secondary structural preferences, local disorder, backbone and side-chain dynamics[69,76,78–80].

Despite the wide range of information that can be obtained, NMR techniques come with limitations, e.g., averaging of fast exchange events and low signal/noise ratio as well as the high amount of protein required and the expensive isotopic labeling[79]. Small Angle X-Ray Scattering (SAXS) allows to investigate at low-resolution (around 20 Å) the geometry of molecules, in particular, it can be applied to study the structure and flexibility of IDPs[81]. On the one hand, the sample preparation is relatively accessible and easy. On the other hand the technique is rather sensitive to inter-particle interactions, potentially distorting the structural parameters extracted when analysing SAXS data, and the apparatus for the measurement is quite complicated[71]. For all these above-mentioned reasons, SAXS is usually combined with NMR techniques or protein crystallography[46,47]. Electron Microscopy (EM) has also been used to visualize and characterize the structure of amyloid fibrils[82–84]. Thanks to the recently solved structure it was possible to identify the different properties between amyloids formed in pathological and physiological conditions[85].

Still concerning the amyloid field, one of the most common and widespread assay to monitor the kinetic of fibrils formation is the use of thioflavin T (ThT) as a fluorescent probe[86,87]. It is based on the change of the emission properties of this fluorophore upon intercalation into β-sheet structures. The assay does not allow a quantification of the fibrils, though[88]. Moreover, ThT is not amyloid specific and it might not be sensitive to monomers or small oligomers, thus it does not sense the early stage of the aggregation[88]. Although intensively used, bulk techniques are not capable of characterizing the properties of amyloid oligomers at a single species level due to their related heterogeneity, metastability, and transient nature, since they provide only average information[89,90]. In this respect, one of the strategies is to stabilize the amyloid oligomers, e.g., by lyophilization, incubation with chemicals, or preparation in membrane-mimicking environments[91–93]. However, these approaches are limited by the incapability of real-time observations[90]. By contrast, Atomic Force Microscopy (AFM) and single-molecule fluorescence techniques have emerged as the most powerful tools in this respect. The first one has been used to acquire 3D morphological maps of biological samples on a support, with sub-nanometric resolution[94–96]. For instance, peak force quantitative nanomechanical mapping (PF-QNM) has been recently implemented to gain insight into the morphology and the mechanical properties at the nanoscale of biological samples[97]. In addition, transient assembly intermediates studies and time-lapse elongation measurements have been performed by using the tapping-mode (TM) both in air and liquid. Air TM AFM was usually preferred because of a better control on the degree of hydration of the sample, often a source of artifacts[98–102]. However, for liquid TM AFM, the sample needs to be well-adhered to the surface, which is the main limitation related to the AFM techniques[103]. To date, a real breakthrough was the development of the technique combining AFM and IR (AFM-IR), used to acquire IR spectra with a 50-100 nm spatial resolution. It enables the collection of spatially resolved IR spectra as well as high-resolution images, acquired at a specific IR wavenumber[104]. Among fluorescence-based techniques, instead, Single-molecule Forster Resonance Energy Transfer (smFRET) allows to characterize oligomer species formed during amyloid fibrillation. In particular, oligomers' size and conformation can be determined by analyzing fluorescence intensity and FRET efficiency of oligomer bursts. However, the samples need to be diluted to nano or picomolar concentration range, which may cause the dissociation of unstable oligomers[90]. Moreover, a covalent label is usually required. A valid alternative is TIRF (Total Internal Reflection Fluorescence) making use of structure-specific dye such as ThT or pentameric formyl thiophene acetic acid (pFTAA), which prevents photobleaching and exogenous effects[90]. The classical resolution (around 200 nm) due to the diffraction limit, has been improved by the development of super-resolution imaging techniques, such as Stochastic Optical Reconstruction Microscopy (STORM) and Stimulated Emission Depletion Microscopy (STED), guaranteeing the investigation of the properties and morphology of the amyloid aggregates at the nanoscale level[105–107]. All of these are crucial aspects when considering that other than a mere conformational point of view, the formation of toxic oligomers and their interaction with cellular membranes is believed to be one possible mechanism of action causing cellular toxicity[108–110]. Thus, it is critical to investigate the molecular mechanism underlying the crucial protein-membrane interaction. For this purpose, most of the biophysical techniques previously mentioned in this chapter have been employed during the last decades[111–113]. However, the possible low population of membrane-bound protein makes those techniques not completely and universally suitable and applicable. Indeed, to quantify the above-mentioned interaction and estimate the dissociation constants ($K_d$s) Surface Plasmon Resonance (SPR), Isothermal Titration Calorimetry (ITC) and Biolayer Interferometry are often used. The main advantage is their sensitivity, whereas their main drawback is that in some cases where more than two species are involved, describing the thermodynamics might be extremely complicated[114]. Furthermore, Quartz Crystal Microbalance with Dissipation Monitoring (QCM-D) is a highly sensitive surface technique often used to monitor frequency change related to the oscillation of the quartz crystal, possibly caused by the interaction of the protein injected into a chamber containing an homogeneous layer of vesicles, e.g., Supported Lipid Bilayer (SLB), LUVs or Small Unilamellar Vesicles (SUVs), previously deposited onto the sensor[115,116]. However, the methods do not provide any information about contacts at the amino acids residue level, in contrast to NMR-based approaches. Specifically, NMR in solution allows to obtain indirect information about the lipid binding, e.g., an $^{1}$H-$^{15}$N HSQC/TROSY (Heteronuclear Single Quantum Coherence/Transverse Relaxation Optimized Spectroscopy) spectrum would show a change in the intensity of some residues rather than a chemical shift, upon lipid-protein interaction[117]. By using Chemical Exchange Saturation Transfer or Dark-State Exchange Saturation Transfer (CEST or DEST) residue-specific saturation transfer profiles can be obtained, giving information about binding sites and affinity. It is based on the drastic decrease of the reorientational motions caused by the binding[118]. However, it is worth underling that to employ this technique big lipid particles must be often substituted by smaller membrane-mimicking systems, such as micelles, bicelles and nanodiscs.

Additionally, the use of high concentrations might often represent an issue[114]. Solid-state NMR has the potential to study structural and dynamic membrane interactions. Some of the advantages related to this technique include the variety in labeling approaches, the sensitivity to the local conformational changes caused by different environmental factors, the possibility to detect both rigid and mobile regions, and the atomic resolution[114]. The dynamics of amyloidogenic proteins interacting with lipid membranes can also be captured by AFM[119–121]. More recently the interaction between Aβ and model membranes has been observed with High-Speed (HS) AFM[122]. NanoIR-AFM and Tip-Enhanced Raman Spectroscopy (TERS) have allowed to investigate the amyloid-membrane/neuronal spine interaction[123,124] whereas Surface Enhanced Raman Spectroscopy (SERS) was able to determine secondary structure at nM concentrations[125]. Fluorescence is one of the most versatile and powerful methods to investigate, directly or indirectly, protein-lipid interactions. One of the ways is to look at the change in the chemico-physical properties of the bilayer as a result of the interaction with the protein of interest, e.g., hydration, polarity, mobility, packing, lateral diffusion or fluidity. The fluorophores commonly used to achieve this goal are grouped according to the position of the moiety with respect to the bilayer. 6-lauroyl-2-dimethylaminonaphthalene (Laurdan), 6-hexadecanoyl2-(((2-(trimethylammonium)ethyl) methyl)amino)naphthalene chloride (Patman)**,** 6-propionyl-2-dimethylaminonaphthalene (Prodan)**,** 4-[(n-dodecylthio)methyl]-7-(N,N-dimethylamino)-coumarin (DTMAC) are only some of them, reported here from deepest to the shallowest[126–128].

However, all of those methods provide indirect evidence of the binding. Fluorescence Correlation Spectroscopy (FCS) or FRET are considered to be more straightforward approaches in this context, but both require a covalent labeling of the protein. FCS relies on the detection of fluorescence intensity fluctuations generated by the emitting molecules going through the focal volume. From that, the autocorrelation curves are obtained and the diffusion coefficient and the concentration calculated[129]. A decrease of the diffusion coefficient compared to the unbound protein (3D diffusion in solution) indicates either a binding with vesicles or aggregation processes. The main disadvantage is that the FCS technique requires a standard and an appropriate calibration. On the contrary, z-scan FCS allows to measure the diffusion time of a lipid or a protein bound to the membrane and, importantly, a standard is not needed[130–132]. In addition, more quantitative measurements can be performed to obtain partition coefficients ($K_p$)[133] and the dissociation constants ($K_d$s) [134]. The latter can be also measured using a Fluorescence Cross-Correlation approach, via labeling vesicles and proteins with different probes[135]. Similarly, flow-cytometric analysis, namely Fluorescence-Activated Cell Sorter (FACS), is a fast and sensitive method which has been optimized to detect vesicles-protein binding[136,137].

FRET is extensively employed for different purposes in biophysics and related disciplines, one of those is to detect protein-lipid interactions[138–141]. It consists of labeling the protein and the vesicles with two different suitable dyes, appropriately chosen so that the emission spectrum of the donor significantly overlaps with the excitation of the acceptor. The lifetime of the donor is thus measured as a function of the acceptor concentration. Due to the occurrence of the FRET, the donor lifetime has to decrease when the two dyes are in close proximity (1-10 nm)[140].

Despite its versatility and sensitivity, fluorescent-labeled proteins usually possess probes not negligible in size, especially if compared with relatively small proteins such as IAPP and Aβ. The presence of such tags might influence the nature and the dynamics of label-free proteins, and for these reasons, label-free techniques should often be combined with fluorescence-based measurements. Also, connecting the early events responsible for the formation of oligomers to the subsequent oligomer-lipid interactions and to the final disruptive consequences might be strongly helpful in a future perspective, e.g., for a proper drug design[142–145]. What has generally been acknowledged is the last stage of the whole process, i.e., cellular death. How does this happen? Is it the oligomer that transfers from the aqueous to the lipid phase, and what is the mechanism? Does the system change its properties? How could this new-formed structure disrupt the bilayer, and why? In this respect, the visualization and the characterization of the intermediates species in the membrane could be helpful, especially in the context of their action mechanism. One of the most common assays used to detect the presence of pores or whether the bilayer is damaged is the dye-leakage assay. It consists of including a fluorescent probe in the vesicle's lumen, where the dye is self-quenched. Once going out the vesicles, due to their poration or rupture, the dye gets diluted in the outside water environment, showing its own emission. Therefore, calcein or carboxyfluorescein, and fluorescent chelating agents (e.g., Fura-2), are commonly used in many labs[114]. However, a characterization of the pores' size or morphology is not possible by using this method. For instance, the lack of dye release might sometimes be due to the wrong choice of the dye and not due to absence of pores[29]. Recently, a fluorescence-based assay has been developed to correlate protein oligomerization and pores in the membrane[146]. Interestingly, one of the few pieces of evidence concerning the amyloids was achieved by AFM, by which a five-subunits pore has been observed in the case of h-IAPP[61]. To conclude, all these examples show which are the methodologies nowadays available and how they could contribute to shed light on different aspects of the same phenomenon and at different time scales, like a photograph taken at different zoom, aperture, and exposure time conditions. However, due to the complexity of the phenomena investigated, the current state of the art allows us to reach a fragmented picture, made of many single essential pieces composing the whole puzzle. Still, further effort is needed to merge the various pieces and achieve a unifying and comprehensive framework.

### 3.2 The simulation dilemma

The terms IDPs or intrinsically disordered regions (IDRs) started appearing because a single 3D structure cannot statistically describe a dynamical ensemble of configurations characterizing such peptides. The configurational flexibility of such molecules means that there is not a single stable structure. However, the environment itself (e.g., ions, buffer, ligand) can stabilize (partially) folded structures.

As a consequence, in recent years this concept gave birth to a parallel database compared to the Protein Data Bank (PDB), which collects experimentally determined ensembles, called Protein Ensemble Database (PED)[147]. Due to the spatial and time resolution limitations of experimental techniques, MD simulations have emerged as an appropriate tool to describe, with atomistic details, the conformational flexibility of peptides. Despite that, the generation of an accurate structure ensemble employing MD simulations is not a straightforward task, and different technical aspects can affect the final output. Firstly, the fuzziness of the peptides is a problem itself, because an accurate sampling of the configurational space requires more simulation time. Secondly, the accuracy of the forcefield (FF) requires a rigorous description of protein-protein, protein-water and water-water interactions. Thirdly, it is hard to define the accuracy of the FF itself, which in turn depends on the available experimental data. Unfortunately, empirical methods that directly probe the configurational ensemble are poor. In most cases, we rely on techniques such as NMR, Small Angle X Ray (SAXS), SANS, CD, FRET, which are powerful but report on the average structure of the molecule in time, losing dynamics information. NMR is in fact the golden standard for the IDPs ensemble determination. Combining different

NMR techniques (see chapter 3.1) we can collect structural information that can be fed in the form of restraint to MD simulations and extract ensembles that are compatible with the experimental data. However, there is no guarantee that the obtained ensemble is the only one compatible with the experiment, i.e., a different ensemble could be representative of the same experimental constraint[79].

Given the intrinsic difficulties of describing IDPs and IDRs, it is not surprising that a lot of effort has been focused on FF development, and some special tweak needed to be done to adjust available FF, originally thought to describe globular or rigid protein, to describe such molecules. We can therefore distinguish between FF originally developed for simulating structured protein and FF developed to simulate IDPs. In the first category we find, for example, some of the most used CHARMM36[148], CHARMM22*[149] and AMBER-ff19SB (AMBER)[150]. In the second category, we have for example CHARMM36m[151], AMBER-ff14IDPSFF[152], and AMBER-ff14IDPs[153]. This list is not exhaustive, considering that the aim of the review is not to describe different FFs. However, it is worth noting that all of these FFs make use of a grid-based energy correction for the $\varphi/\psi$ peptide angle 2D potential energy surface (PES), commonly known as CMAP correction[154]. This allows a better fit of $\varphi$ and $\psi$ angle distribution, therefore the secondary structure propensity. Of course, the correction will depend on the experimental training set used to estimate the angles propensity.

Given that different FFs exist and that the PES heavily depend on the parametrization strategy, it does not surprise that using different FFs might give different results. As a consequence, there is no general consensus on which FF gives the most accurate outcome. A few examples concerning the variability of the results and the dependence on the set of experimental data employed will be reported here.

CHARMM36 is an improvement of CHARMM22* for folded protein, however, Rauscher et al.[155] showed that CHARMM22* was more accurate on reproducing IDPs ensembles for the data set that they have used. They also stressed how critical is the choice of the initial structure to generate the ensemble.

Later, CHARMM36m is an improvement of CHARMM36, aimed at simulating IDPs. Zapletal V. et al.[156] found that the combination of CHARMM36m with TIP4P-D performs better than CHARMM22* according to their experimental data set, while Rahman et al.[157] found that CHARMM22* performs better than CHARMM36m on average using a different data set. They also proved, comparing six FFs, that AMBER-ff14IDPs and AMBER-ff14IDPSFF averagely score better than others, but not for all the proteins used in the data set; on a positive note, they found that in most of the cases under study IDP-specific FFs agree better with the experiments.

They further showed that the results strongly depend on the initial structure and this feature is common for all IDP-specific FFs.

In our opinion, two approaches can be adopted to solve, at least, some of the problems aforementioned. The first approach includes methods of enhanced sampling such as Metadynamics[158] or (Temperature, Hamiltonian) Replica exchange MD[159] to overcome the limitation of choosing a specific single initial structure and sample the configuration space as much as we can, to generate a more realistic and representative ensemble. These methods are nowadays easy to implement and very flexible. They are, however, more suitable for simple systems, such as a single protein in water or protein dimerization. When the system or the phenomenon under study is more complicated, i.e., the simultaneous binding/folding of an IDP with a target protein/molecule, the choice of a collective variable to bias the simulation could be very difficult and the convergence of the simulation might not be reached in a reasonable time scale. The second approach is to use information obtained experimentally to drive our simulation. This is a quite broad approach. For example, we can use Bayesian statistics to generate an experimental-driven ensemble starting from FFs's generated configurations[160]. In cases where the IDPs assume a more defined structure, we could either use experimental information to choose the starting configuration (and even restrain the configuration during the simulations) or we choose the FF better describing the experimental configuration[145]. In our opinion, this means that the choice of the FF and the starting configurations are the first sources of simulation bias.

Due to the chameleonic nature of amyloidogenic proteins and the fast and complex biological processes dynamics, a bottom-up approach is needed to better understand the experiments on living cell experiments. Indeed, *in vitro* and computational methods are usually employed to reduce the degree of complexity of the systems to study. However, for the above mentioned limitations, neither *in vitro* experiments nor computational techniques used separately can give a comprehensive overview of the membrane damage mechanisms caused by amyloidogenic proteins, due to their spatial and temporal resolution limits. Therefore, we believe that only a combination of the methods could provide a comprehensive overview. This belief drove us to the choice of the approach when demonstrating and verifying the lipid-chaperone hypothesis.

**4.0 Lipid-chaperone hypothesis**

The framework behind the lipid-chaperon hypothesis relies on the thermodynamic partitioning of lipids between the aqueous phase and membrane-like organelles (bilayers, micelles, etc.). Thermodynamics states that there will always be a concentration of free lipids at equilibrium in the aqueous phase, in dynamical exchange with the lipid phase. This concentration is defined as the Critical Micellar Concentration (CMC). Our initial hypothesis was that these free lipids in solutions could interact with amyloidogenic proteins in the water phase, giving rise to a lipid-protein complex, globally possessing a higher hydrophobicity and, in turn, a higher affinity to the membrane compared to the bare protein see **Figure 1 A** for a schematic representation).

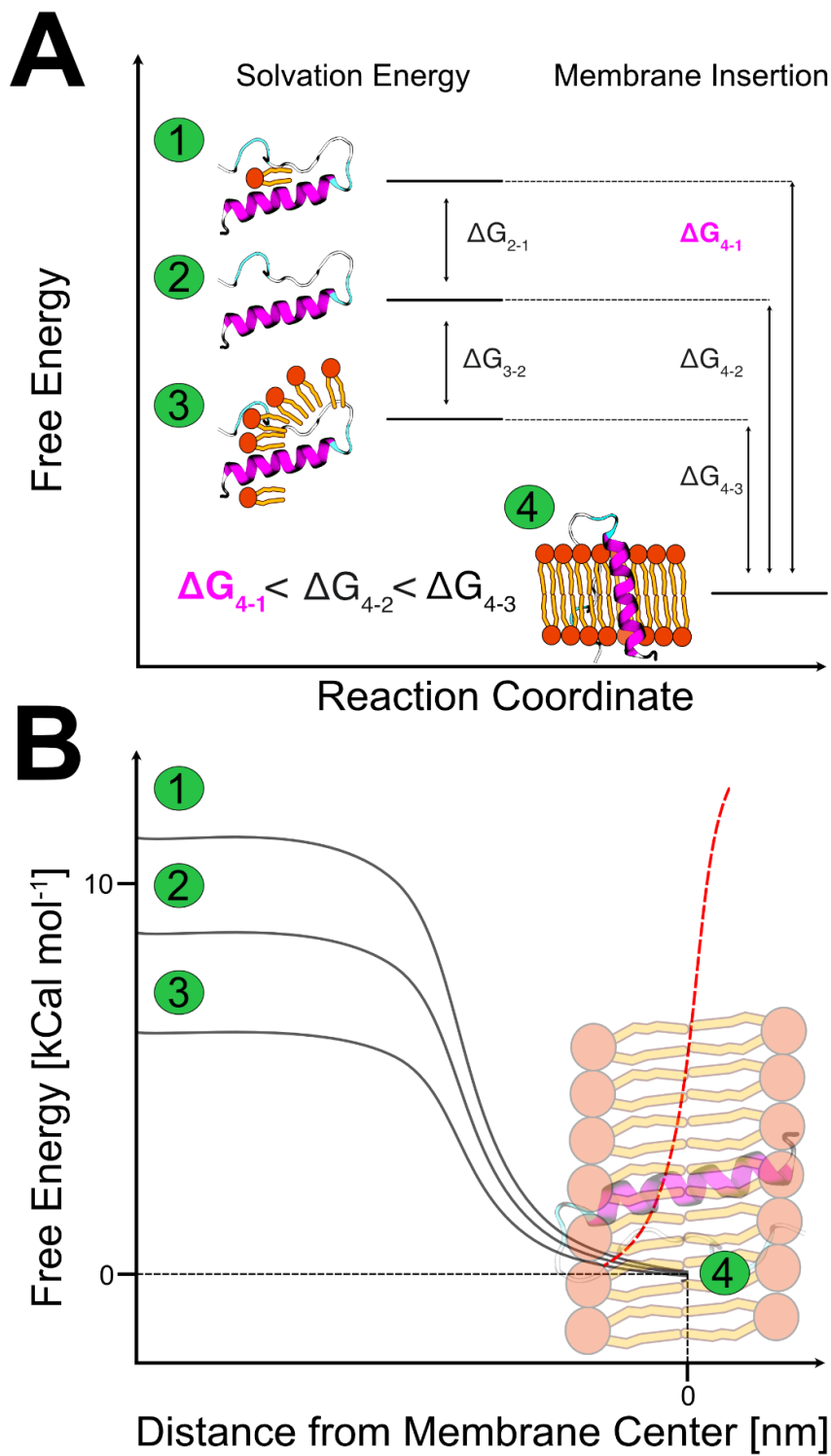

**Figure 1.** (A) Schematic representations of the theoretical model. 1, 2, and 3 represent three different (initial) states of the protein in water; 4 represent the (final) state of protein embedded into the membrane. $\Delta G_{j-i}$ is the free energy between the final state (4) and initial state (1,2 or 3). (B) Schematic representation of Free energy profiles calculated with biased molecular dynamics simulations for amyloidogenic IDPs - membrane insertion. Both model and MD calculations are performed for the bare protein and in complex with n lipids

(n=1 and n>3). The dotted red curve represents the different behavior observed between MD simulations and the theoretical model.

Therefore, we proposed a diffusion-reaction model that could describe the kinetic effect of lipid-assisted transport, such as the protein insertion into a bilayer[26]. The results of the model is a parametric equation which can be fed with crucial parameters and output the protein coverage of the membrane as a function of time. The key parameters are the following: **CMC**, water/membrane partition coefficient of the protein (**P**), binding association constant for protein-lipid complex (**$K_{eq}$**), insertion rate (**D**) of both the bare protein and the protein-lipid complex. The estimation of these parameters is the crucial step of the model. The entry rate **D** of the bare protein or the protein-lipid complex depends on the hydrophobicity: the higher the hydrophobicity, the higher the entry rate. Following this assumption, the kinetic model allows us to calculate the protein insertion grade (**γ**) on the membrane as a function of the CMC and the protein concentration (if the other parameters are estimated).

Mainly three important considerations have to be made:

1. The effect of the lipid-assisted transport is significant at high protein-lipid complex concentration, which in turn requires high **$K_{eq}$**.
2. The complex should be more hydrophobic and have a higher translocation rate into the bilayer than the bare protein.
3. The equilibrium constant for the complex formation should be high.

MD simulations confirmed the latter (3), showing that the complex formation is irreversible[26].

However, gaining insight into the translocation rate **D** would require a deep knowledge about the mechanism of insertion, which is very challenging from either an experimental or a computational point of view. The proposed model asserts that the difference in the entry rate **D** between the bare protein and protein-lipid complex is given by the solvation energy in water (**Figure 1 B**), which can be estimated using experiment or MD simulations regardless of the entry mechanism.

The last point to be discussed is the stoichiometry of the complex. The number of lipids binding to the protein can be any. Few considerations, however, have to be drawn in this respect. Firstly, the concentration of free lipids in water is usually a few orders of magnitude lower than the protein's. We, therefore, assume that the most likely stoichiometry would be 1:1 from a statistical point of view. Secondly, the protein could act as a nucleation site for the lipids, forming a micelle-like structure when the number of lipids bound to the protein increases. This kind of structure would likely not have a higher hydrophobicity than the bare protein because of the amphiphilic nature of the lipids, as illustrated in **Figure 1 A-B**.

Given all of these considerations, we believed it was reasonable to hypothesize a 1:1 protein-lipid complex. Starting from this model, we employed a combined experimental and computational approach to corroborate our hypothesis. The first step was to verify that the protein-lipid complex's hydrophobicity is higher than the bare protein. Therefore, we calculate the free energy of insertion for the bare protein, the 1:1 (protein:lipid) complex, and 1:3 complex in a POPC bilayer using biased molecular dynamic simulations (**Figure 1 B**)[28].

Few differences compared to the original paper are reported here. First, in the illustration of **Figure 1 B**, the profiles are aligned to the minimum of the free energy profile corresponding to a distance where the protein is inserted into the bilayer. Instead, in the original paper, they are aligned to the water phase, where the proteins (and the lipid-protein complexes) are at a non-interacting distance from the membrane surface. Even though this difference is non changing the meaning of the results, we realized that the original choice of aligning the plots to their state in water is not consistent with the diffusion-reaction model. Based on the diffusion-reaction model, indeed, since the hydrophobicity of the bare protein is different from the lipid-protein complex, they should not have the same free energy when in water ( see $\Delta G_{2-1}$ and $\Delta G_{3-2}$ in **Figure 1 A)**.

Furthermore, the diffusion-reaction model assumes that the final state (membrane-inserted state, **Figure 1 A**, point 4) is equal for all systems (bare protein, protein-lipid complexes 1:1, and 1:3). Another difference to the original free energy profiles of the F. Scollo and C. Tempra et al.[28] work, is the increase in the free energy upon the membrane absorption (**Figure 1 B**, red dotted line). The free energy increases due to the initial protein configuration used for the calculation, which was mainly non-structured (random coil). After the absorption on the membrane, the systems were not simulated long enough to catch the secondary structure transition to alpha-helix, which would have favored the membrane insertion. As described in chapter 3, the choice of the initial configuration is biasing the results. With hindsight, we should have used a protein pre-structured in alpha-helix as a starting point for the free energy calculations since our later developments of the characterization of the protein-lipid complex show, experimentally, a formation of alpha-helix structure upon the binding with the lipid[29] (see **Figure 2 A-B**).

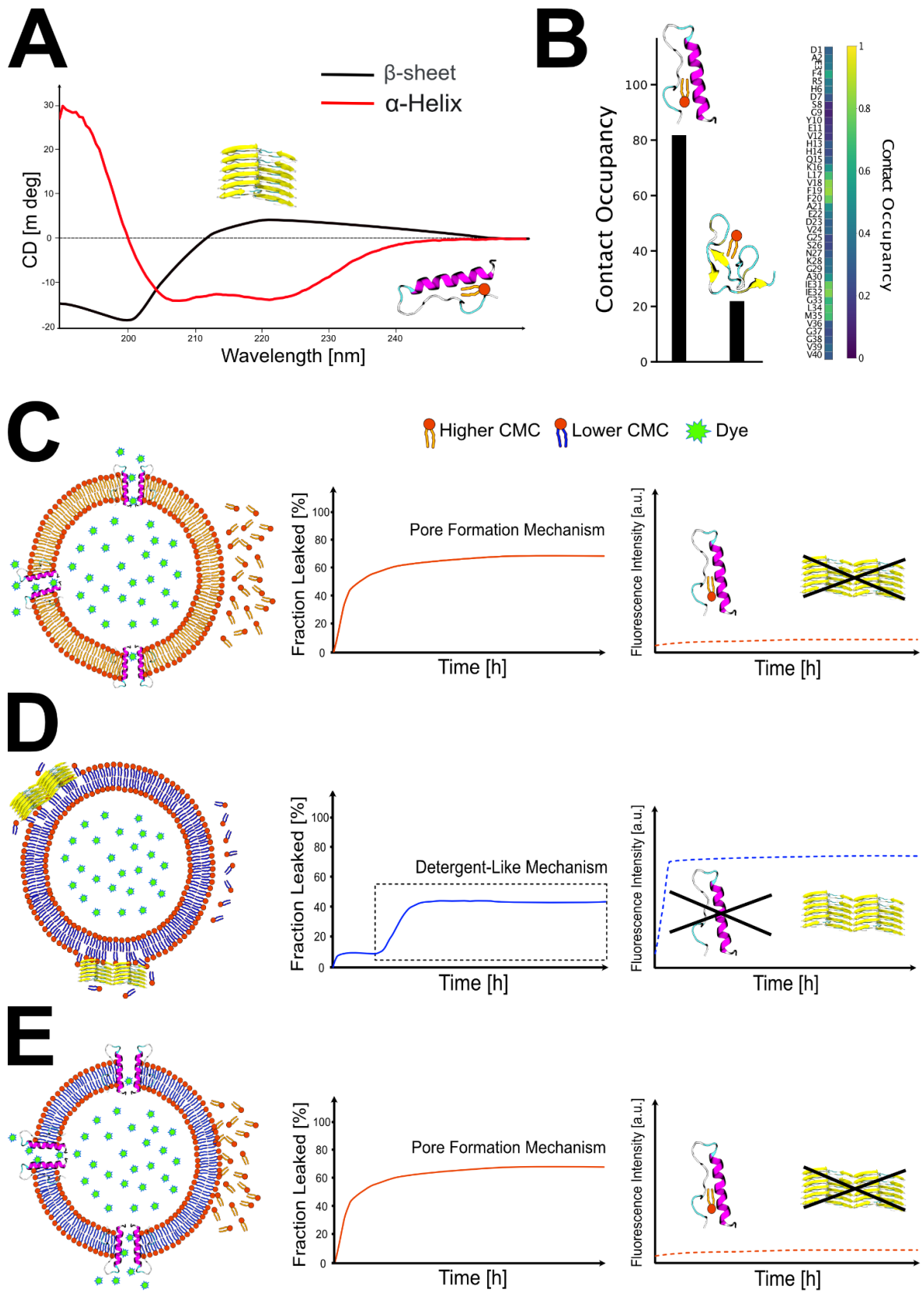

**Figure 2. A)** Schematic representation of Circular Dichroism Spectra showing α-helix and β-sheet structures, red and blue curves, respectively associated to a generic proteins in the presence of free lipids and to the protein in the absence of lipids undergoing fibrillation. **B)** Contact occupancy of the 1:1 lipid-protein complexes. The bar plot shows a higher contact occupancy when the protein secondary structure is α-helix,

compared to a β-sheet structure. Panel **C**, **D** and **E** show a schematic representation of the three different conditions studied and published in[28]. The examples can be extended to the data in[29]. **C)** In the first case the vesicles used are the shorter chain, thus the higher CMC. Here the formation of pores is favoured (red curve) and no fibrillation (red dotted curve). **D)** for the longest chain phospholipids (low CMC), the detergent-like mechanism is promoted (blue curve) and fibrils content is higher (blue dotted curve). **E)** The thicker vesicles (lower CMC) in the presence of the free lipids (higher CMC) are characterized by the same trend observed in panel **C**, i.e., pore formation and no fibrils (solid and dotted red curves, respectively).

For these reasons, we reported in **Figure 1 B** an alpha-helix structure protein instead of the random coil one used in our previous publication.

From an experimental point of view, one of the ways to modulate the concentration of free lipids, always at the equilibrium with the vesicles, is to change the CMC of the systems under study. Also, to avoid possible artifacts due to the drastic change of the bilayer properties, we decided to minimize the differences between different vesicles. Thus homogeneous LUVs composed of the same polar head (phosphatidyl-coline – PC) but different chain lengths (reflecting a different CMC) lipids have been employed to study the effect of the lipid CMC on amylin aggregation and its subsequent interaction with model membranes. Intuitively, the longer the hydrophobic tail, the smaller is the CMC. More specifically, five different diacyl-phosphocholine phospholipids (spanning from 14 to 22 carbon atoms and with a unique unsaturation -PC14, PC16, PC18, PC20, PC22) have been chosen after determining their CMC by using pyrene fluorescence[161]. H-IAPP was used as a model amyloidogenic protein. Under the same conditions of protein and vesicles concentrations, pH, and ionic strength, ThT and dye-leakage assays have been performed. ThT assay has shown that fibrils were not detected in the presence of the shorter phospholipids, i.e., higher CMC, as reported in **Figure 2 C**.

The data indicate that higher concentrations of free lipids inhibit fibril growth, whereas lower concentrations promote this process (**Figure 2 C-D**). Interestingly, the same occurred when studying the amylin fibrillation in the absence of vesicles by adding the protein to the supernatant obtained upon centrifugation of the LUVs. The outcome of the experiments was justified by assuming that the lipid-protein interactions are more favored when the concentration of free lipids is higher. In addition, this interaction prevents the protein-protein one, leading to the formation of fibrils as the ultimate stage. Under the same conditions, carboxy-fluorescein was enclosed upon extrusion into the lumen of the vesicles, and the dye-release was monitored. The obtained curve could be interpreted using the so-called two-step mechanism[25], as depicted in **Figure 2D**. According to that, membrane damage can occur through two steps, as better described in chapter 2. The first is the formation of tiny pores, and the second is the removal of phospholipids, i.e., via a detergent-like mechanism. Our data demonstrate that at higher CMCs, when the lipid-protein complex is more likely to form, the first step is enhanced and the second one repressed (**Figure 2 C-D**). In other words, the hypothesized lipid-protein complex might penetrate the membrane via pore formation. Furthermore, a critical experiment was performed to exclude the potential role of the thickness of the bilayer (hydrophobic mismatch - see **Figure 2 E**). Indeed, decreasing the CMC by increasing the number of carbon atoms composing the hydrophobic tail affects proportionally the thickness of the bilayer, which could have been the reason why amylin is not capable of forming pores through the thicker bilayer. The aforementioned experiment consisted of diluting the thicker vesicles (PC22) in a solution containing the shortest free phospholipid (PC14). Interestingly, no fibrils were detected for this system. Moreover, the curve obtained for the dye-leakage experiments was comparably similar to the one previously obtained for the PC14 vesicles, as illustrated in **Figure 2 E**. Thanks to these experiments, possible artifacts caused by the different thicknesses of the bilayer were excluded. For a better description and illustration of the data, we address the reader to our previous work[28]. In this work, we used a combined experimental and computational approach to show, to the best of our knowledge for the first time, that the free lipids in dynamic equilibrium with the LUVs mediate the amylin insertion. Specifically, CMC acts as a switch between pore formation, membrane disruption, and fibril growth. However, the experimental techniques did not directly detect the lipid-protein interaction in the water phase. Therefore, the following driving questions have been aroused: a) can we directly prove the existence of the lipid-protein complex? b) could this new model be extended to other amyloidogenic proteins? Could it be considered as a mechanism shared by all or at least some of them?

Moved by these questions, another set of experiments was carried out. Specifically, we studied the behavior of Aβ $_{(1-40)}$, αS, βS, and r-IAPP, as amyloidogenic and non-amyloidogenic proteins, respectively[29]. Firstly, in the latter work, the behavior of the amyloidogenic Aβ and αS was compared to the previous one shown for the h-IAPP. The same experiments were carried out under the same conditions, and these have shown that the free phospholipids affect the behaviour of these proteins toward the aggregation and the interaction with the bilayer. Indeed, analogously to the previously described protein, Aβ and αS aggregation is repressed by a higher concentration of free lipids, which by contrast promotes pores formation, although with a different kinetic. It is worth noting that the pores formed by Aβ have been found to be significantly smaller compared to the h-IAPP and αS; for this reason, the Fura-2 assay needed to be used. Proven that the CMC is a discriminating factor in enhancing or inhibiting either pore formation or detergent-like mechanism and fibril formation, we studied with the same approach two non-amyloidogenic proteins, the r-IAPP, and βS. The relation between the ability to induce pores in model membranes and the cytotoxicity has been proven to be highly complex. Our work has shown that in the presence of any lipid compositions used, the r-IAPP cannot form fibrils, in agreement with all the data reported in the literature. By contrast, with the highest amount of free phospholipids, the dye-release was observed for the r-IAPP as well.

Furthermore, βS was compared to αS and r-IAPP, and the dye-leakage kinetic was monitored using the thickest LUVs (PC22) diluted in free PC14, the same system represented in **Figure 2E**. Unlike the other proteins, βS does not damage the membrane, not even in the presence of a higher concentration of free lipid. This is probably due to the lack of a hydrophobic region, which is thought to be responsible for the aggregation propensity. Moreover, we employed CD, 2D NMR and MD simulations to answer the first question, identify and characterize the protein-lipid complex. Aβ $_{(1-40)}$ has been structurally characterized in the absence and the presence of 1,2-dimyristoyl-sn-glycero-3-phosphocholine (DMPC 14:0). A significant chemical shift change after 1 hour of incubation has been observed through 2D NMR measurements, indicating mainly the occurrence of a hydrophobic interaction. These results were corroborated by time-lapse CD spectra, showing that in the absence of lipids, the classical random coil to β-sheet transition was observed, while in the presence of free PC14:0, a random-coil to α-helix conformational change was reported (**Figure 2 A**).

A further set of MD simulations was carried out to gain atomistic details about the protein-lipid complex and the amino acids responsible for the interaction, as reported in **Figure 2 B**. It is crucial to notice that the CD experiment of Aβ $_{(1-40)}$ in the absence and presence of lipids in solutions at their CMC showed a considerable fraction of alpha-helix structure in the presence of lipids (**Figure 2 A**). Due to the current limitation of FFs to reproduce exact ensemble conformation (see chapter 3), we decided to employ MD simulations of the different proteins starting from a pre-formed alpha-helical structure. Our analysis showed that the hydrophobic tail of the lipid strongly interacts with the alpha-helical region of the protein (**Figure 2 B**).

Furthermore, when comparing Aβ $_{(1-40)}$ and Aβ $_{(1-42)}$, or h-IAPP and r-IAPP, we found a positive correlation between alpha-helix content and binding. Among all proteins, r-IAPP had a minor helical structure content, either at the beginning of the simulations and after its binding to lipid. r-IAPP was as well the protein with the weaker binding with the lipid.

Our combined experimental and computational approach has been fundamental to demonstrate the conformational transition's crucial role to alpha-helix upon binding to free-lipids in the aqueous phase. The preformed protein-lipid complex in water and its alpha-helix structure increase the hydrophobicity of the protein, fostering membrane absorption and insertion[148], as also suggested by H. Fatafta et al.[151]. These new findings correlate well with our previous model, suggesting that the helical structure is responsible for h-IAPP and Aβ oligomerization inside the membrane and pore formation [145,163–165].
To summarize, the lipid-chaperone hypothesis addresses the hydrophobicity as the driving force of the possible formation of the lipid-protein complex, and therefore the bilayer penetration is affected, as a consequence of that.

## 5.0 Conclusions and New Directions

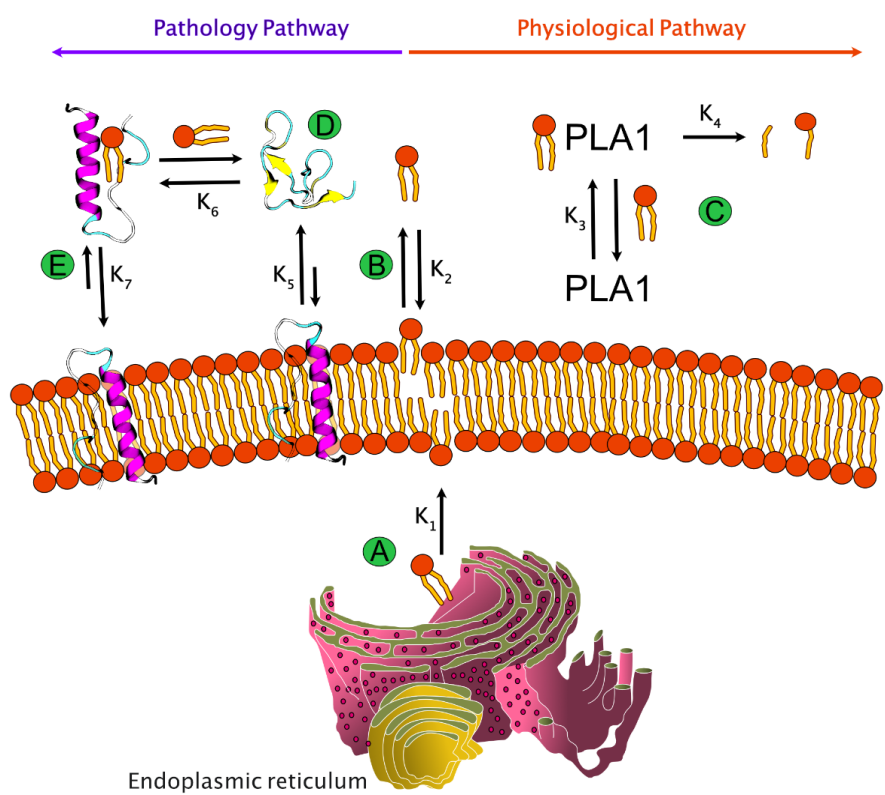

**Figure 3.** Schematic representation of the lipid-chaperone hypothesis in a biological context.

The lipid-chaperone hypothesis is based on the presence of free lipids in the aqueous phase, forming a stable complex with the amyloidogenic protein, which is then transported into the bilayer. This idea is supported by *in vitro* experiments on amyloidogenic proteins such as h-IAPP, Aβ, and αS, and non-amyloidogenic r-IAPP and βS, taken as control[29]. From these data, two relevant questions arise: What are the sources of free lipids? Does the lipid-chaperone hypothesis assume a biological relevance?

Several cellular defense mechanisms exist to prevent undesired effects deriving from protein aggregation. For instance, chaperons can shield exposed hydrophobic patches of oligomers and prevent them to stick to cellular components or further assemble into fibers. With aging, chaperones levels decrease significantly (~30%), whereas the serum concentration of esther-linked phosphatidylcholine (PC) and phosphatidylethanolamine (PE) increases, as does their chance to interact with solvent-exposed hydrophobic domains of amyloidogenic peptides. Furthermore, an abnormal lipid metabolism is observed in individuals developing T2D, AD, and PD can contribute to the onset of these pathologies[31–34,166]. In physiological conditions, phospholipids synthesis occurs mainly in the endoplasmic reticulum, as shown in **Figure 3 A**. They are transported by chaperones and lipid transport proteins (LTP) into the bilayer to replace phospholipids going from the bilayer to the aqueous phase, as chemical equilibrium requires (**Figure 3 B**). These phospholipids in the aqueous phase are captured by the phospholipase A1 enzyme and transformed into lysophospholipids and fatty acids (**Figure 3 C**)[167]. This is the mechanism by which free phospholipids are removed from the aqueous phase. Perhaps in physiological conditions, this phospholipids path works in steady-state conditions as shown in **Figure 3**. In other words, the reaction rates of in and out lipid transport are equal. In pathological conditions, the equilibrium in **Figure 3 B** might be more shifted towards the formation of free lipids due to a miss function of phospholipase enzymes. Then, amyloidogenic IDPs could sequester free lipids forming a stable complex. Similarly, another source of free lipids could also be due to oxidized phospholipids. Indeed, in physiological conditions, reactive oxygen species (ROSs) are continually formed during molecular oxygen exchange and destroyed by oxidant machinery, e.g., superoxide dismutase-catalase enzyme. Under pathological conditions, instead, redox homeostasis is unbalanced, and an excess of lipid peroxides is produced[168]. It should be noted that lipid peroxides have a high CMC (about μM)[169] compared to the non oxidized phospholipids (about 0.01 μM)[170] and they have already been shown to play a crucial role in type II diabetes, Alzheimer's, and Parkinson's pathologies[168,171–179]. The chemical equilibrium law suggests that lipid-protein complexes can be formed by increasing protein or free lipid concentrations. In physiological conditions, amyloidogenic proteins have been shown to play a positive modulatory effect in the picomolar range, whereas higher nanomolar concentrations lead to detrimental effects that can culminate in dementia, as demonstrated in the case of Aβ[180]. As previously proposed, the source of increased free lipids could be due to oxidized phospholipids[29] or to a mis-functioning of phospholipase. Thus, we can speculate that an increase of free lipids or peptide concentration fosters lipid-protein complex formation. In turn, the hydrophobicity of the complex, compared to the protein and lipid alone, increases, as does its propensity to transfer from the aqueous phase to the membrane, as schematized in **Figure 3 D-E**. The lipid-chaperone hypothesis is a general molecular model including both amyloid and toxic oligomers hypotheses. The amyloid hypothesis addresses fibrils as the leading actors in membrane damage. In contrast, the toxic oligomer hypothesis proposes that small unstructured oligomers form ion-channel-like pores responsible for membrane damage. The lipid-chaperone hypothesis suggests that the concentration of free phospholipid in the aqueous phase acts as a switch for pore or fibril formation, i.e., low lipid concentration favors fibrils, whereas pores form at high lipid concentration. At intermediate concentration, both fibrils and pores are formed.

In light of the last findings, the role of free phospholipids should be taken into account in the experimental data interpretation, drug and diagnostic development.

Recently, the Food and Drug Administration (FDA) has approved a monoclonal antibody named aducanumab (Aduhelm) by Abiogen Pharma as a drug against Alzheimer's, massively disappointing the international scientific community[181]. Aduhelm's development is based on the amyloid hypothesis: its action is based on desegregating fibrils deposits in the brain. In the news reported by Nature's paper, the amyloid hypothesis received many criticisms from a neuroscientist because it is too reductive to explain a multifaceted disease such as Alzheimer's, Parkinson or diabetes.

In the effective drugs development and instrumental diagnostic, the role of free lipid should be considered and further investigated. We strongly believe that the lipid-chaperone hypothesis may open new strategies and routes to be studied to understand better and fight amyloid-related diseases.

# Acknowledgement

All the authors are grateful to Professor Antonio Raudino, who passed away in 2021, for envisioning and contributing to the development of the Lipid-chaperone.

C.T. would like to acknowledge the International Max Planck Research School for Many-Particle Systems in Structured Environments hosted by the Max Planck Institute for the Physics of Complex Systems, Dresden, Germany. C.T. thanks the HPC-Europa3 Transna- tional Access program (GA #730897, application HPC174QJYP). F.S. would like to acknowledge the support by the Czech Science Foundation GA CR EXPRO (grant 19-26854X) and thank the HPC-Europa 3 Translational Access program (GA #730897 application HPC17MRQUB). F.L. gratefully acknowledges the data storage service SDS@hd supported by the Ministry of Science, Research and the Arts Baden-Württemberg (MWK) and the German Research Foundation (DFG) through grant INST 35/1314-1 FUGG and INST 35/1503-1 FUGG.

**References**

1. Uversky, V. N. Intrinsically Disordered Proteins and Their "Mysterious" (Meta)Physics. *Front. Phys.* **7**, 10 (2019).

2. Drake, J. A. & Pettitt, B. M. Physical Chemistry of the Protein Backbone: Enabling the Mechanisms of Intrinsic Protein Disorder. *J. Phys. Chem. B* **124**, 4379–4390 (2020).

3. Lella, M. & Mahalakshmi, R. Metamorphic Proteins: Emergence of Dual Protein Folds from One Primary Sequence. *Biochemistry* **56**, 2971–2984 (2017).

4. Bahramali, G., Goliaei, B., Minuchehr, Z. & Salari, A. Chameleon sequences in neurodegenerative diseases. *Biochem. Biophys. Res. Commun.* **472**, 209–216 (2016).

5. Wright, P. E. & Dyson, H. J. Intrinsically disordered proteins in cellular signalling and regulation. *Nat. Rev. Mol. Cell Biol.* **16**, 18–29 (2015).

6. Milardi, D., Pappalardo, M., Pannuzzo, M., Grasso, D. M. & Rosa, C. L. The role of the Cys2-Cys7 disulfide bridge in the early steps of Islet amyloid polypeptide aggregation: A molecular dynamics study. *Chem. Phys. Lett.* **463**, 396–399 (2008).

7. Milardi, D. *et al.* Proteostasis of Islet Amyloid Polypeptide: A Molecular Perspective of Risk Factors and Protective Strategies for Type II Diabetes. *Chem. Rev.* **121**, 1845–1893 (2021).

8. Nguyen, P. H. *et al.* Amyloid Oligomers: A Joint Experimental/Computational Perspective on Alzheimer's Disease, Parkinson's Disease, Type II Diabetes, and Amyotrophic Lateral Sclerosis. *Chem. Rev.* **121**, 2545–2647 (2021).

9. Rauk, A. The chemistry of Alzheimer's disease. *Chem. Soc. Rev.* **38**, 2698 (2009).

10. Pannuzzo, M. Beta‑amyloid pore linked to controlled calcium influx into the cell: A new paradigm for Alzheimer's Disease. *Alzheimers Dement.* alz.12373 (2021) doi:10.1002/alz.12373.

11. García-Viñuales, S. *et al.* The interplay between lipid and Aβ amyloid homeostasis in Alzheimer's Disease: risk factors and therapeutic opportunities. *Chem. Phys. Lipids* **236**, 105072 (2021).

12. Cirrito, J. R. *et al.* Synaptic Activity Regulates Interstitial Fluid Amyloid-β Levels In Vivo. *Neuron* **48**, 913–922 (2005).

13. Kamenetz, F. *et al.* APP Processing and Synaptic Function. *Neuron* **37**, 925–937 (2003).

14. Maries, E., Dass, B., Collier, T. J., Kordower, J. H. & Steece-Collier, K. The role of α-synuclein in Parkinson's disease: insights from animal models. *Nat. Rev. Neurosci.* **4**, 727–738 (2003).

15. Meade, R. M., Fairlie, D. P. & Mason, J. M. Alpha-synuclein structure and Parkinson's disease – lessons and emerging principles. *Mol. Neurodegener.* **14**, 29 (2019).

16. Sengupta, U., Nilson, A. N. & Kayed, R. The Role of Amyloid-β Oligomers in Toxicity, Propagation, and Immunotherapy. *EBioMedicine* **6**, 42–49 (2016).

17. Silver, B. L. *The physical chemistry of membranes: an introduction to the structure and dynamics of biological membranes*. (2013).

18. Pappalardo, M., Milardi, D., Grasso, D. & La Rosa, C. Phase behaviour of polymer-grafted DPPC membranes for drug delivery systems design. *J. Therm. Anal. Calorim.* **80**, 413–418 (2005).

19. Raudino, Antonio., Zuccarello, Felice., La Rosa, Carmelo. & Buemi, Giuseppe. Thermal expansion and compressibility coefficients of phospholipid vesicles: experimental determination and theoretical modeling. *J. Phys. Chem.* **94**, 4217–4223 (1990).

20. Chiti, F. & Dobson, C. M. Protein Misfolding, Amyloid Formation, and Human Disease: A Summary of Progress Over the Last Decade. *Annu. Rev. Biochem.* **86**,

27–68 (2017).

21. Selkoe, D. J. & Hardy, J. The amyloid hypothesis of Alzheimer's disease at 25 years. *EMBO Mol. Med.* **8**, 595–608 (2016).

22. La Rosa, C. *et al.* Symmetry-breaking transitions in the early steps of protein self-assembly. *Eur. Biophys. J.* **49**, 175–191 (2020).

23. Brender, J. R., Salamekh, S. & Ramamoorthy, A. Membrane Disruption and Early Events in the Aggregation of the Diabetes Related Peptide IAPP from a Molecular Perspective. *Acc. Chem. Res.* **45**, 454–462 (2012).

24. Scalisi, S. *et al.* Self-Assembling Pathway of HiApp Fibrils within Lipid Bilayers. *ChemBioChem* **11**, 1856–1859 (2010).

25. Sciacca, M. F. M. *et al.* Two-Step Mechanism of Membrane Disruption by Aβ through Membrane Fragmentation and Pore Formation. *Biophys. J.* **103**, 702–710 (2012).

26. La Rosa, C., Scalisi, S., Lolicato, F., Pannuzzo, M. & Raudino, A. Lipid-assisted protein transport: A diffusion-reaction model supported by kinetic experiments and molecular dynamics simulations. *J. Chem. Phys.* **144**, 184901 (2016).

27. Korshavn, K. J. *et al.* Reduced Lipid Bilayer Thickness Regulates the Aggregation and Cytotoxicity of Amyloid-β. *J. Biol. Chem.* **292**, 4638–4650 (2017).

28. Scollo, F. *et al.* Phospholipids Critical Micellar Concentrations Trigger Different Mechanisms of Intrinsically Disordered Proteins Interaction with Model Membranes. *J. Phys. Chem. Lett.* **9**, 5125–5129 (2018).

29. Sciacca, M. F. *et al.* Lipid-Chaperone Hypothesis: A Common Molecular Mechanism of Membrane Disruption by Intrinsically Disordered Proteins. *ACS Chem. Neurosci.* **11**, 4336–4350 (2020).

30. Gallardo, J., Escalona-Noguero, C. & Sot, B. Role of α-Synuclein Regions in Nucleation and Elongation of Amyloid Fiber Assembly. *ACS Chem. Neurosci.* **11**, 872–879 (2020).

31. Bisaglia, M. *et al.* Structure and topology of the non-amyloid-β component fragment of human α-synuclein bound to micelles: Implications for the aggregation process. *Protein Sci.* **15**, 1408–1416 (2006).

32. Di Scala, C. *et al.* Common molecular mechanism of amyloid pore formation by Alzheimer's β-amyloid peptide and α-synuclein. *Sci. Rep.* **6**, 28781 (2016).

33. Giasson, B. I., Murray, I. V. J., Trojanowski, J. Q. & Lee, V. M.-Y. A Hydrophobic Stretch of 12 Amino Acid Residues in the Middle of α-Synuclein Is Essential for Filament Assembly. *J. Biol. Chem.* **276**, 2380–2386 (2001).

34. Sahoo, B. R. *et al.* A cationic polymethacrylate-copolymer acts as an agonist for β-amyloid and an antagonist for amylin fibrillation. *Chem. Sci.* **10**, 3976–3986 (2019).

35. Glenner, G. G. & Wong, C. W. Alzheimer's disease: Initial report of the purification and characterization of a novel cerebrovascular amyloid protein. *Biochem. Biophys. Res. Commun.* **120**, 885–890 (1984).

36. Goate, A. *et al.* Segregation of a missense mutation in the amyloid precursor protein gene with familial Alzheimer's disease. *Nature* **349**, 704–706 (1991).

37. Rogaev, E. I. *et al.* Familial Alzheimer's disease in kindreds with missense mutations in a gene on chromosome 1 related to the Alzheimer's disease type 3 gene. *Nature* **376**, 775–778 (1995).

38. Sherrington, R. *et al.* Cloning of a gene bearing missense mutations in early-onset familial Alzheimer's disease. *Nature* **375**, 754–760 (1995).

39. Westermark, P. Aspects on human amyloid forms and their fibril polypeptides: Human amyloid forms and their fibril polypeptides. *FEBS J.* **272**, 5942–5949 (2005).

40. Rambaran, R. N. & Serpell, L. C. Amyloid fibrils: Abnormal protein assembly. *Prion* **2**, 112–117 (2008).

41. Lorenzo, A. & Yankner, B. A. Amyloid Fibril Toxicity in Alzheimer's Disease and Diabetesa. *Ann. N. Y. Acad. Sci.* **777**, 89–95 (1996).

42. Stéphan, A., Laroche, S. & Davis, S. Generation of Aggregated β-Amyloid in the Rat Hippocampus Impairs Synaptic Transmission and Plasticity and Causes Memory Deficits. *J. Neurosci.* **21**, 5703–5714 (2001).

43. Pieri, L., Madiona, K., Bousset, L. & Melki, R. Fibrillar α-Synuclein and Huntingtin Exon 1 Assemblies Are Toxic to the Cells. *Biophys. J.* **102**, 2894–2905 (2012).

44. Okada, T., Wakabayashi, M., Ikeda, K. & Matsuzaki, K. Formation of Toxic Fibrils of Alzheimer's Amyloid β-Protein-(1–40) by Monosialoganglioside GM1, a Neuronal Membrane Component. *J. Mol. Biol.* **371**, 481–489 (2007).

45. Reynolds, N. P. *et al.* Mechanism of Membrane Interaction and Disruption by α-Synuclein. *J. Am. Chem. Soc.* **133**, 19366–19375 (2011).

46. Kakio, A., Nishimoto, S., Yanagisawa, K., Kozutsumi, Y. & Matsuzaki, K. Interactions of Amyloid β-Protein with Various Gangliosides in Raft-Like Membranes: Importance of GM1 Ganglioside-Bound Form as an Endogenous Seed for Alzheimer Amyloid. *Biochemistry* **41**, 7385–7390 (2002).

47. Milanesi, L. *et al.* Direct three-dimensional visualization of membrane disruption by amyloid fibrils. *Proc. Natl. Acad. Sci.* **109**, 20455–20460 (2012).

48. Erten-Lyons, D. *et al.* Factors associated with resistance to dementia despite high Alzheimer disease pathology. *Neurology* **72**, 354–360 (2009).

49. Sloane, J. A. *et al.* Lack of correlation between plaque burden and cognition in the aged monkey. *Acta Neuropathol. (Berl.)* **94**, 471–478 (1997).

50. Jack, C. R. *et al.* Age-specific population frequencies of cerebral β-amyloidosis and neurodegeneration among people with normal cognitive function aged 50–89 years: a cross-sectional study. *Lancet Neurol.* **13**, 997–1005 (2014).

51. McLean, C. A. *et al.* Soluble pool of Aβ amyloid as a determinant of severity of neurodegeneration in Alzheimer's disease. *Ann. Neurol.* **46**, 860–866 (1999).

52. *Diabetes: clinical science in practice*. (Cambridge University Press, 1995).

53. Scollo, F. & La Rosa, C. Amyloidogenic Intrinsically Disordered Proteins: New Insights into Their Self-Assembly and Their Interaction with Membranes. *Life* **10**, 144 (2020).

54. Westermark, P., Andersson, A. & Westermark, G. T. Islet Amyloid Polypeptide, Islet Amyloid, and Diabetes Mellitus. *Physiol. Rev.* **91**, 795–826 (2011).

55. Meier, J. J. *et al.* Inhibition of human IAPP fibril formation does not prevent β-cell death: evidence for distinct actions of oligomers and fibrils of human IAPP. *Am. J. Physiol.-Endocrinol. Metab.* **291**, E1317–E1324 (2006).

56. Konarkowska, B., Aitken, J. F., Kistler, J., Zhang, S. & Cooper, G. J. S. The aggregation potential of human amylin determines its cytotoxicity towards islet β-cells. *FEBS J.* **273**, 3614–3624 (2006).

57. Haataja, L., Gurlo, T., Huang, C. J. & Butler, P. C. Islet Amyloid in Type 2 Diabetes, and the Toxic Oligomer Hypothesis. *Endocr. Rev.* **29**, 303–316 (2008).

58. Ono, K. The Oligomer Hypothesis in α-Synucleinopathy. *Neurochem. Res.* **42**, 3362–3371 (2017).

59. Hebda, J. A. & Miranker, A. D. The Interplay of Catalysis and Toxicity by Amyloid Intermediates on Lipid Bilayers: Insights from Type II Diabetes. *Annu. Rev. Biophys.* **38**, 125–152 (2009).

60. Kourie, J. I., Culverson, A. L., Farrelly, P. V., Henry, C. L. & Laohachai, K. N. Heterogeneous Amyloid-Formed Ion Channels as a Common Cytotoxic Mechanism. *Cell Biochem. Biophys.* **36**, 191–207 (2002).

61. Quist, A. *et al.* Amyloid ion channels: A common structural link for protein-misfolding disease. *Proc. Natl. Acad. Sci.* **102**, 10427–10432 (2005).

62. Zhao, J. *et al.* Probing ion channel activity of human islet amyloid polypeptide (amylin). *Biochim. Biophys. Acta BBA - Biomembr.* **1818**, 3121–3130 (2012).

63. Olzscha, H. *et al.* Amyloid-like Aggregates Sequester Numerous Metastable Proteins

with Essential Cellular Functions. *Cell* **144**, 67–78 (2011).

64. Stöckl, M. T., Zijlstra, N. & Subramaniam, V. α-Synuclein Oligomers: an Amyloid Pore?: Insights into Mechanisms of α-Synuclein Oligomer–Lipid Interactions. *Mol. Neurobiol.* **47**, 613–621 (2013).

65. Li, S. *et al.* Soluble A Oligomers Inhibit Long-Term Potentiation through a Mechanism Involving Excessive Activation of Extrasynaptic NR2B-Containing NMDA Receptors. *J. Neurosci.* **31**, 6627–6638 (2011).

66. Juhola, H. *et al.* Negatively Charged Gangliosides Promote Membrane Association of Amphipathic Neurotransmitters. *Neuroscience* **384**, 214–223 (2018).

67. Click, T. H., Ganguly, D. & Chen, J. Intrinsically Disordered Proteins in a Physics-Based World. *Int. J. Mol. Sci.* **11**, 5292–5309 (2010).

68. Clarke, D. T. Circular Dichroism in Protein Folding Studies. *Curr. Protoc. Protein Sci.* **70**, (2012).

69. Dunker, A. K. *et al.* Intrinsically disordered protein. *J. Mol. Graph. Model.* **19**, 26–59 (2001).

70. Sarroukh, R., Goormaghtigh, E., Ruysschaert, J.-M. & Raussens, V. ATR-FTIR: A "rejuvenated" tool to investigate amyloid proteins. *Biochim. Biophys. Acta BBA - Biomembr.* **1828**, 2328–2338 (2013).

71. Schramm, A. *et al.* An arsenal of methods for the experimental characterization of intrinsically disordered proteins – How to choose and combine them? *Arch. Biochem. Biophys.* **676**, 108055 (2019).

72. Moran, S. D. & Zanni, M. T. How to Get Insight into Amyloid Structure and Formation from Infrared Spectroscopy. *J. Phys. Chem. Lett.* **5**, 1984–1993 (2014).

73. Flynn, J. D., Jiang, Z. & Lee, J. C. Segmental $^{13}$C‑Labeling and Raman Microspectroscopy of α‑Synuclein Amyloid Formation. *Angew. Chem. Int. Ed.* **57**, 17069–17072 (2018).

74. Li, H., Lantz, R. & Du, D. Vibrational Approach to the Dynamics and Structure of Protein Amyloids. *Molecules* **24**, 186 (2019).

75. Watson, M. D. & Lee, J. C. Coupling chemical biology and vibrational spectroscopy for studies of amyloids in vitro and in cells. *Curr. Opin. Chem. Biol.* **64**, 90–97 (2021).

76. Eliezer, D. Biophysical characterization of intrinsically disordered proteins. *Curr. Opin. Struct. Biol.* **19**, 23–30 (2009).

77. Habchi, J. *et al.* Monitoring Structural Transitions in IDPs by Site-Directed Spin Labeling EPR Spectroscopy. in *Intrinsically Disordered Protein Analysis* (eds. Uversky, V. N. & Dunker, A. K.) vol. 895 361–386 (Humana Press, 2012).

78. Gibbs, E. B. & Showalter, S. A. Quantitative Biophysical Characterization of Intrinsically Disordered Proteins. *Biochemistry* **54**, 1314–1326 (2015).

79. Habchi, J., Tompa, P., Longhi, S. & Uversky, V. N. Introducing Protein Intrinsic Disorder. *Chem. Rev.* **114**, 6561–6588 (2014).

80. Konrat, R. NMR contributions to structural dynamics studies of intrinsically disordered proteins. *J. Magn. Reson.* **241**, 74–85 (2014).

81. Kikhney, A. G. & Svergun, D. I. A practical guide to small angle X-ray scattering (SAXS) of flexible and intrinsically disordered proteins. *FEBS Lett.* **589**, 2570–2577 (2015).

82. Guerrero-Ferreira, R. *et al.* Cryo-EM structure of alpha-synuclein fibrils. *eLife* **7**, e36402 (2018).

83. Han, S. *et al.* Amyloid plaque structure and cell surface interactions of β-amyloid fibrils revealed by electron tomography. *Sci. Rep.* **7**, 43577 (2017).

84. Röder, C. *et al.* Cryo-EM structure of islet amyloid polypeptide fibrils reveals similarities with amyloid-β fibrils. *Nat. Struct. Mol. Biol.* **27**, 660–667 (2020).

85. Ragonis-Bachar, P. & Landau, M. Functional and pathological amyloid structures in the eyes of 2020 cryo-EM. *Curr. Opin. Struct. Biol.* **68**, 184–193 (2021).

86. Xue, C., Lin, T. Y., Chang, D. & Guo, Z. Thioflavin T as an amyloid dye: fibril quantification, optimal concentration and effect on aggregation. *R. Soc. Open Sci.* **4**, 160696 (2017).

87. Sciacca, M. F. M., Monaco, I., La Rosa, C. & Milardi, D. The active role of $Ca^{2+}$ ions in Aβ-mediated membrane damage. *Chem. Commun.* **54**, 3629–3631 (2018).

88. LeVine, H. Quantification of β-sheet amyloid fibril structures with thioflavin T. in *Methods in Enzymology* vol. 309 274–284 (Elsevier, 1999).

89. Simone Ruggeri, F., Habchi, J., Cerreta, A. & Dietler, G. AFM-Based Single Molecule Techniques: Unraveling the Amyloid Pathogenic Species. *Curr. Pharm. Des.* **22**, 3950–3970 (2016).

90. Yang, J., Perrett, S. & Wu, S. Single Molecule Characterization of Amyloid Oligomers. *Molecules* **26**, 948 (2021).

91. Chen, S. W. *et al.* Structural characterization of toxic oligomers that are kinetically trapped during α-synuclein fibril formation. *Proc. Natl. Acad. Sci.* **112**, E1994–E2003 (2015).

92. Ciudad, S. *et al.* Aβ(1-42) tetramer and octamer structures reveal edge conductivity pores as a mechanism for membrane damage. *Nat. Commun.* **11**, 3014 (2020).

93. Ehrnhoefer, D. E. *et al.* EGCG redirects amyloidogenic polypeptides into unstructured, off-pathway oligomers. *Nat. Struct. Mol. Biol.* **15**, 558–566 (2008).

94. Adamcik, J. & Mezzenga, R. Study of amyloid fibrils via atomic force microscopy. *Curr. Opin. Colloid Interface Sci.* **17**, 369–376 (2012).

95. Martorana, V. *et al.* Amyloid jams: Mechanical and dynamical properties of an amyloid fibrillar network. *Biophys. Chem.* **253**, 106231 (2019).

96. Sokolov, D. V. Atomic Force Microscopy for Protein Nanotechnology. in *Protein Nanotechnology* (ed. Gerrard, J. A.) vol. 996 323–367 (Humana Press, 2013).

97. Zhang, S., Aslan, H., Besenbacher, F. & Dong, M. Quantitative biomolecular imaging

by dynamic nanomechanical mapping. *Chem Soc Rev* **43**, 7412–7429 (2014).

98. Ding, T. T., Lee, S.-J., Rochet, J.-C. & Lansbury, P. T. Annular α-Synuclein Protofibrils Are Produced When Spherical Protofibrils Are Incubated in Solution or Bound to Brain-Derived Membranes. *Biochemistry* **41**, 10209–10217 (2002).

99. Kad, N. M. *et al.* Hierarchical Assembly of β2-Microglobulin Amyloid In Vitro Revealed by Atomic Force Microscopy. *J. Mol. Biol.* **330**, 785–797 (2003).

100. Zhu, M., Han, S., Zhou, F., Carter, S. A. & Fink, A. L. Annular Oligomeric Amyloid Intermediates Observed by in Situ Atomic Force Microscopy. *J. Biol. Chem.* **279**, 24452–24459 (2004).

101. Chromy, B. A. *et al.* Self-Assembly of Aβ$_{1-42}$ into Globular Neurotoxins. *Biochemistry* **42**, 12749–12760 (2003).

102. Griffo, A. *et al.* Single-Molecule Force Spectroscopy Study on Modular Resilin Fusion Protein. *ACS Omega* **2**, 6906–6915 (2017).

103. Gosal, W., Myers, S., Radford, S. & Thomson, N. Amyloid Under the Atomic Force Microscope. *Protein Pept. Lett.* **13**, 261–270 (2006).

104. Dazzi, A. *et al.* AFM–IR: Combining Atomic Force Microscopy and Infrared Spectroscopy for Nanoscale Chemical Characterization. *Appl. Spectrosc.* **66**, 1365–1384 (2012).

105. Lee, J.-E. *et al.* Mapping Surface Hydrophobicity of α-Synuclein Oligomers at the Nanoscale. *Nano Lett.* **18**, 7494–7501 (2018).

106. Spehar, K. *et al.* Super‐resolution Imaging of Amyloid Structures over Extended Times by Using Transient Binding of Single Thioflavin T Molecules. *ChemBioChem* **19**, 1944–1948 (2018).

107. Torra, J., Bondia, P., Gutierrez-Erlandsson, S., Sot, B. & Flors, C. Long-term STED imaging of amyloid fibers with exchangeable Thioflavin T. *Nanoscale* **12**, 15050–15053 (2020).

108. Bemporad, F. & Chiti, F. Protein Misfolded Oligomers: Experimental Approaches, Mechanism of Formation, and Structure-Toxicity Relationships. *Chem. Biol.* **19**, 315–327 (2012).

109. Cleary, J. P. *et al.* Natural oligomers of the amyloid-β protein specifically disrupt cognitive function. *Nat. Neurosci.* **8**, 79–84 (2005).

110. Winner, B. *et al.* In vivo demonstration that -synuclein oligomers are toxic. *Proc. Natl. Acad. Sci.* **108**, 4194–4199 (2011).

111. de Planque, M. R. R. *et al.* β-Sheet Structured β-Amyloid(1-40) Perturbs Phosphatidylcholine Model Membranes. *J. Mol. Biol.* **368**, 982–997 (2007).

112. Divakara, M. B. *et al.* Molecular mechanisms for the destabilization of model membranes by islet amyloid polypeptide. *Biophys. Chem.* **245**, 34–40 (2019).

113. Terzi, E., Hölzemann, G. & Seelig, J. Interaction of Alzheimer β-Amyloid Peptide(1−40) with Lipid Membranes. *Biochemistry* **36**, 14845–14852 (1997).

114. Chang, Z., Deng, J., Zhao, W. & Yang, J. Exploring interactions between lipids and amyloid-forming proteins: A review on applying fluorescence and NMR techniques. *Chem. Phys. Lipids* **236**, 105062 (2021).

115. Cuco, A., Serro, A. P., Farinha, J. P., Saramago, B. & da Silva, A. G. Interaction of the Alzheimer Aβ(25–35) peptide segment with model membranes. *Colloids Surf. B Biointerfaces* **141**, 10–18 (2016).

116. Griffo, A. *et al.* Design and Testing of a Bending‑Resistant Transparent Nanocoating for Optoacoustic Cochlear Implants. *ChemistryOpen* **8**, 1100–1108 (2019).

117. Bodner, C. R., Dobson, C. M. & Bax, A. Multiple Tight Phospholipid-Binding Modes of α-Synuclein Revealed by Solution NMR Spectroscopy. *J. Mol. Biol.* **390**, 775–790 (2009).

118. Fawzi, N. L., Ying, J., Ghirlando, R., Torchia, D. A. & Clore, G. M. Atomic-resolution dynamics on the surface of amyloid-β protofibrils probed by

solution NMR. *Nature* **480**, 268–272 (2011).

119. Pilkington, A. W. *et al.* Acetylation of Aβ$_{40}$ Alters Aggregation in the Presence and Absence of Lipid Membranes. *ACS Chem. Neurosci.* **11**, 146–161 (2020).

120. Kinoshita, M. *et al.* Model membrane size-dependent amyloidogenesis of Alzheimer's amyloid-β peptides. *Phys. Chem. Chem. Phys.* **19**, 16257–16266 (2017).

121. Sciacca, M. F. M. *et al.* The Role of Cholesterol in Driving IAPP-Membrane Interactions. *Biophys. J.* **111**, 140–151 (2016).

122. Ewald, M. *et al.* High speed atomic force microscopy to investigate the interactions between toxic Aβ$_{1-42}$ peptides and model membranes in real time: impact of the membrane composition. *Nanoscale* **11**, 7229–7238 (2019).

123. Feuillie, C. *et al.* High Speed AFM and NanoInfrared Spectroscopy Investigation of Aβ1–42 Peptide Variants and Their Interaction With POPC/SM/Chol/GM1 Model Membranes. *Front. Mol. Biosci.* **7**, 571696 (2020).

124. Tabatabaei, M., Caetano, F. A., Pashee, F., Ferguson, S. S. G. & Lagugné-Labarthet, F. Tip-enhanced Raman spectroscopy of amyloid β at neuronal spines. *The Analyst* **142**, 4415–4421 (2017).

125. D'Urso, L. *et al.* Detection and characterization at nM concentration of oligomers formed by hIAPP, Aβ(1–40) and their equimolar mixture using SERS and MD simulations. *Phys. Chem. Chem. Phys.* **20**, 20588–20596 (2018).

126. Amaro, M. *et al.* Time-Resolved Fluorescence in Lipid Bilayers: Selected Applications and Advantages over Steady State. *Biophys. J.* **107**, 2751–2760 (2014).

127. Scollo, F. *et al.* What Does Time-Dependent Fluorescence Shift (TDFS) in Biomembranes (and Proteins) Report on? *Front. Chem.* **9**, 738350 (2021).

128. Jurkiewicz, P., Sýkora, J., Olżyńska, A., Humpolíčková, J. & Hof, M. Solvent Relaxation in Phospholipid Bilayers: Principles and Recent Applications. *J. Fluoresc.* **15**, 883–894 (2005).

129. Eigen, M. & Rigler, R. Sorting single molecules: application to diagnostics and evolutionary biotechnology. *Proc. Natl. Acad. Sci.* **91**, 5740–5747 (1994).

130. Amaro, M. *et al.* GM$_1$ Ganglioside Inhibits β‑Amyloid Oligomerization Induced by Sphingomyelin. *Angew. Chem. Int. Ed.* **55**, 9411–9415 (2016).

131. Humpolíčková, J. *et al.* Probing Diffusion Laws within Cellular Membranes by Z-Scan Fluorescence Correlation Spectroscopy. *Biophys. J.* **91**, L23–L25 (2006).

132. Štefl, M., Kułakowska, A. & Hof, M. Simultaneous Characterization of Lateral Lipid and Prothrombin Diffusion Coefficients by Z-Scan Fluorescence Correlation Spectroscopy. *Biophys. J.* **97**, L1–L3 (2009).

133. Melo, A. M., Prieto, M. & Coutinho, A. The effect of variable liposome brightness on quantifying lipid–protein interactions using fluorescence correlation spectroscopy. *Biochim. Biophys. Acta BBA - Biomembr.* **1808**, 2559–2568 (2011).

134. Rusu, L., Gambhir, A., McLaughlin, S. & Rädler, J. Fluorescence Correlation Spectroscopy Studies of Peptide and Protein Binding to Phospholipid Vesicles. *Biophys. J.* **87**, 1044–1053 (2004).

135. Krüger, D., Ebenhan, J., Werner, S. & Bacia, K. Measuring Protein Binding to Lipid Vesicles by Fluorescence Cross-Correlation Spectroscopy. *Biophys. J.* **113**, 1311–1320 (2017).

136. Temmerman, K. & Nickel, W. A novel flow cytometric assay to quantify interactions between proteins and membrane lipids. *J. Lipid Res.* **50**, 1245–1254 (2009).

137. Lolicato, F. *et al. Cholesterol promotes both head group visibility and clustering of PI(4,5)P$_2$ driving unconventional secretion of Fibroblast Growth Factor 2.* http://biorxiv.org/lookup/doi/10.1101/2021.04.16.440132 (2021) doi:10.1101/2021.04.16.440132.

138. Coutinho, A., Loura, L. M. S. & Prieto, M. FRET studies of lipid-protein aggregates related to amyloid-like fibers: FRET studies of amyloid-like aggregates. *J.*

*Neurochem.* **116**, 696–701 (2011).

139. Kakio, A., Nishimoto, S., Kozutsumi, Y. & Matsuzaki, K. Formation of a membrane-active form of amyloid β-protein in raft-like model membranes. *Biochem. Biophys. Res. Commun.* **303**, 514–518 (2003).

140. Munishkina, L. A. & Fink, A. L. Fluorescence as a method to reveal structures and membrane-interactions of amyloidogenic proteins. *Biochim. Biophys. Acta BBA - Biomembr.* **1768**, 1862–1885 (2007).

141. Wong, P. T. *et al.* Amyloid-β Membrane Binding and Permeabilization are Distinct Processes Influenced Separately by Membrane Charge and Fluidity. *J. Mol. Biol.* **386**, 81–96 (2009).

142. Sciacca, Michele. F. M. *et al.* Inhibition of Aβ Amyloid Growth and Toxicity by Silybins: The Crucial Role of Stereochemistry. *ACS Chem. Neurosci.* **8**, 1767–1778 (2017).

143. Lolicato, F., Raudino, A., Milardi, D. & La Rosa, C. Resveratrol interferes with the aggregation of membrane-bound human-IAPP: A molecular dynamics study. *Eur. J. Med. Chem.* **92**, 876–881 (2015).

144. Romanucci, V. *et al.* Modulating Aβ aggregation by tyrosol-based ligands: The crucial role of the catechol moiety. *Biophys. Chem.* **265**, 106434 (2020).

145. Tempra, C., La Rosa, C. & Lolicato, F. The role of alpha-helix on the structure-targeting drug design of amyloidogenic proteins. *Chem. Phys. Lipids* **236**, 105061 (2021).

146. Šachl, R. *et al.* Functional Assay to Correlate Protein Oligomerization States with Membrane Pore Formation. *Anal. Chem.* **92**, 14861–14866 (2020).

147. Lazar, T. *et al.* PED in 2021: a major update of the protein ensemble database for intrinsically disordered proteins. *Nucleic Acids Res.* **49**, D404–D411 (2021).

148. Best, R. B. *et al.* Optimization of the Additive CHARMM All-Atom Protein Force

Field Targeting Improved Sampling of the Backbone ϕ, ψ and Side-Chain $\chi_1$ and $\chi_2$ Dihedral Angles. *J. Chem. Theory Comput.* **8**, 3257–3273 (2012).

149. Mackerell, A. D., Feig, M. & Brooks, C. L. Extending the treatment of backbone energetics in protein force fields: Limitations of gas-phase quantum mechanics in reproducing protein conformational distributions in molecular dynamics simulations. *J. Comput. Chem.* **25**, 1400–1415 (2004).

150. Tian, C. *et al.* ff19SB: Amino-Acid-Specific Protein Backbone Parameters Trained against Quantum Mechanics Energy Surfaces in Solution. *J. Chem. Theory Comput.* **16**, 528–552 (2020).

151. Huang, J. *et al.* CHARMM36m: an improved force field for folded and intrinsically disordered proteins. *Nat. Methods* **14**, 71–73 (2017).

152. Bunney, P. E., Zink, A. N., Holm, A. A., Billington, C. J. & Kotz, C. M. Orexin activation counteracts decreases in nonexercise activity thermogenesis (NEAT) caused by high-fat diet. *Physiol. Behav.* **176**, 139–148 (2017).

153. Wang, W., Ye, W., Jiang, C., Luo, R. & Chen, H. New Force Field on Modeling Intrinsically Disordered Proteins. *Chem. Biol. Drug Des.* **84**, 253–269 (2014).

154. MacKerell, A. D., Feig, M. & Brooks, C. L. Improved Treatment of the Protein Backbone in Empirical Force Fields. *J. Am. Chem. Soc.* **126**, 698–699 (2004).

155. Rauscher, S. *et al.* Structural Ensembles of Intrinsically Disordered Proteins Depend Strongly on Force Field: A Comparison to Experiment. *J. Chem. Theory Comput.* **11**, 5513–5524 (2015).

156. Zapletal, V. *et al.* Choice of Force Field for Proteins Containing Structured and Intrinsically Disordered Regions. *Biophys. J.* **118**, 1621–1633 (2020).

157. Rahman, M. U., Rehman, A. U., Liu, H. & Chen, H.-F. Comparison and Evaluation of Force Fields for Intrinsically Disordered Proteins. *J. Chem. Inf. Model.* **60**, 4912–4923 (2020).

158. Laio, A. & Parrinello, M. Escaping free-energy minima. *Proc. Natl. Acad. Sci.* **99**, 12562–12566 (2002).

159. Sugita, Y. & Okamoto, Y. Replica-exchange molecular dynamics method for protein folding. *Chem. Phys. Lett.* **314**, 141–151 (1999).

160. Fisher, C. K., Huang, A. & Stultz, C. M. Modeling Intrinsically Disordered Proteins with Bayesian Statistics. *J. Am. Chem. Soc.* **132**, 14919–14927 (2010).

161. Aguiar, J., Carpena, P., Molina-Bolívar, J. A. & Carnero Ruiz, C. On the determination of the critical micelle concentration by the pyrene 1:3 ratio method. *J. Colloid Interface Sci.* **258**, 116–122 (2003).

162. Fatafta, H., Kav, B., Bundschuh, B. F., Loschwitz, J. & Strodel, B. Disorder-to-order transition of the amyloid-β peptide upon lipid binding. *Biophys. Chem.* **280**, 106700 (2022).

163. Pannuzzo, M. On the physiological/pathological link between Aβ peptide, cholesterol, calcium ions and membrane deformation: A molecular dynamics study. *Biochim. Biophys. Acta BBA - Biomembr.* **1858**, 1380–1389 (2016).

164. Pannuzzo, M., Raudino, A., Milardi, D., La Rosa, C. & Karttunen, M. α-Helical Structures Drive Early Stages of Self-Assembly of Amyloidogenic Amyloid Polypeptide Aggregate Formation in Membranes. *Sci. Rep.* **3**, 2781 (2013).

165. Pannuzzo, M., Raudino, A. & Böckmann, R. A. Peptide-induced membrane curvature in edge-stabilized open bilayers: A theoretical and molecular dynamics study. *J. Chem. Phys.* **141**, 024901 (2014).

166. Cao, P. *et al.* Islet amyloid polypeptide toxicity and membrane interactions. *Proc. Natl. Acad. Sci.* **110**, 19279–19284 (2013).

167. Richmond, G. S. & Smith, T. K. Phospholipases A1. *Int. J. Mol. Sci.* **12**, 588–612 (2011).

168. Ahmad, W., Ijaz, B., Shabbiri, K., Ahmed, F. & Rehman, S. Oxidative toxicity in

diabetes and Alzheimer's disease: mechanisms behind ROS/ RNS generation. *J. Biomed. Sci.* **24**, 76 (2017).

169. Pande, A. H., Kar, S. & Tripathy, R. K. Oxidatively modified fatty acyl chain determines physicochemical properties of aggregates of oxidized phospholipids. *v* **1798**, 442–452 (2010).

170. Marsh, D. *CRC handbook of lipid bilayers*. (CRC Press, 1990).

171. Reale, M., Brenner, T., Greig, N. H., Inestrosa, N. & Paleacu, D. Neuroinflammation, AD, and Dementia. *Int. J. Alzheimers Dis.* **2010**, 1–2 (2010).

172. Asmat, U., Abad, K. & Ismail, K. Diabetes mellitus and oxidative stress—A concise review. *Saudi Pharm. J.* **24**, 547–553 (2016).

173. Gunn, A. P. *et al.* Amyloid-β Peptide Aβ3pE-42 Induces Lipid Peroxidation, Membrane Permeabilization, and Calcium Influx in Neurons. *J. Biol. Chem.* **291**, 6134–6145 (2016).

174. Wei, Z., Li, X., Li, X., Liu, Q. & Cheng, Y. Oxidative Stress in Parkinson's Disease: A Systematic Review and Meta-Analysis. *Front. Mol. Neurosci.* **11**, 236 (2018).

175. Wright, E., Scism-Bacon, J. L. & Glass, L. C. Oxidative stress in type 2 diabetes: the role of fasting and postprandial glycaemia: Oxidative Stress in Type 2 Diabetes. *Int. J. Clin. Pract.* **60**, 308–314 (2006).

176. Lyras, L., Cairns, N. J., Jenner, A., Jenner, P. & Halliwell, B. An Assessment of Oxidative Damage to Proteins, Lipids, and DNA in Brain from Patients with Alzheimer's Disease. *J. Neurochem.* **68**, 2061–2069 (2002).

177. Mattson, M. P., Pedersen, W. A., Duan, W., Culmsee, C. & Camandola, S. Cellular and Molecular Mechanisms Underlying Perturbed Energy Metabolism and Neuronal Degeneration in Alzheimer's and Parkinson's Diseases. *Ann. N. Y. Acad. Sci.* **893**, 154–175 (1999).

178. Pilkington, A. W. *et al.* Hydrogen Peroxide Modifies Aβ–Membrane Interactions with

Implications for Aβ$_{40}$ Aggregation. *Biochemistry* **58**, 2893–2905 (2019).

179. Puspita, L., Chung, S. Y. & Shim, J. Oxidative stress and cellular pathologies in Parkinson's disease. *Mol. Brain* **10**, 53 (2017).

180. Puzzo, D. *et al.* Picomolar Amyloid- Positively Modulates Synaptic Plasticity and Memory in Hippocampus. *J. Neurosci.* **28**, 14537–14545 (2008).

181. Mullard, A. Controversial Alzheimer's drug approval could affect other diseases. *Nature* **595**, 162–163 (2021).